\documentclass[%
 aip,
 amsmath,amssymb,
 reprint,%
]{revtex4-1}

\usepackage{graphicx}
\usepackage{dcolumn}
\usepackage{bm}

\usepackage[utf8]{inputenc}
\usepackage[T1]{fontenc}
\usepackage{mathptmx}

\begin{document}

\preprint{AIP/123-QED}

\title{Non-contact Thermal Transistor Effects Modulated by Nanoscale Mechanical Deformation}

\author{Fangqi Chen}
\affiliation{Department of Mechanical and Industrial Engineering, Northeastern University, Boston, MA, 02115, USA}
\author{Xiaojie Liu}
\affiliation{Department of Mechanical and Industrial Engineering, Northeastern University, Boston, MA, 02115, USA}
\author{Yanpei Tian}
\affiliation{Department of Mechanical and Industrial Engineering, Northeastern University, Boston, MA, 02115, USA}
\author{Duanyang Wang}
\affiliation{Department of Physics, University of California, Santa Barbara, Santa Barbara, CA, 93106, USA}
\author{Yi Zheng}

 \email{y.zheng@northeastern.edu.}
 
\affiliation{Department of Mechanical and Industrial Engineering, Northeastern University, Boston, MA, 02115, USA}
\affiliation{Department of Electrical and Computer Engineering, Northeastern University, Boston, MA, 02115, USA}

%
%
\date{\today}

\begin{abstract}
Thermal management has become a promising field in recent years due to the limitation of energy resources and the global warming. An important topic in improving the efficiency of thermal energy utilization is how to control the flows of heat, and thermal rectifiers, such as the thermal transistor, have been proposed as units for modulating the flow of heat. In this work, a reconfigurable non-contact thermal transistor with two-dimensional grating is introduced. The thermal transistor consists of three parts: source, gate, and drain, with the gate working around the phase-transition temperature of vanadium dioxide, a type of phase-transition material. Results show that the unit has a clear transistor-like behavior. The surface phonon/plasmon polaritons supported by the insulating/metallic states that contribute to the radiative thermal transport can be modulated at a nanoscale separation. And the dynamic amplification factor ranges from 15.4 to 30.6 when the stretchable polydimethylsiloxane is subjected to tension or compression. This work sheds light on studies about the controllable small-scale thermal transport due to mechanical deformations.
\end{abstract}

\maketitle
The electric diode and transistor have been widely used in microelectronics, and they are the cornerstone components for almost all the devices in this field. The diode is a rectifier that exhibits a high degree of asymmetry in the current when the applied voltage bias is reversed. The transistor has a function of modulating the current across the source and drain terminals by controlling the current and temperature of the gate. On the other hand, thermal management has become an important and challenging research field due to the limitation of energy resources on the earth and the approaching energy issues like global warming. So, analogous to electronic diode and transistor, thermal diode and transistor have drawn a lot of attention in the recent years \cite{yang2013radiation,ben2014near,ghanekar2016high,ito2014experimental,joulain2015modulation,prod2016optimized} because of their ability to control the heat flux effectively.

The thermal transistor is a device with the idea brought from the metal-oxide-semiconductor field-effect transistor (MOSFET) and it has three terminals: source, drain, and gate. It can control the heat flux across the source and drain terminals by a little amount of heat added to or removed from the gate terminal. The thermal transistor has three main functions: thermal switch, thermal modulation, and thermal amplification. \cite{ben2014near} While the transistor working as a switch, a small variation of the gate temperature will lead to a large reduction of heat flux lost by the source and received by the drain by a large magnitude, which can be regarded as a switch with "on" and "off" modes. The transistor can also set the flux among certain magnitudes in a small temperature range of the gate. As the most promising function for the thermal transistor, the amplification effect has been widely explored in several studies. Li \emph{et al.} proposed a phonon-based thermal transistor to control heat flow as a counterpart of electric transistor for the first time. \cite{li2006negative} Several studies have been conducted on the far-field \cite{joulain2015modulation,prod2016optimized} and near-field \cite{ben2014near,ghanekar2018near} thermal transistors. The amplification factor is a vital parameter to evaluate the performance of the transistor, defined as $\alpha=|\partial Q_D/\partial Q_G|$, representing how large the heat flux $Q_D$ received by the drain changes due to the change in the heat flux $Q_G$ interacting with the gate terminal. Prod’homme \emph{et al.} carried out a research about the optimization of the thermal amplification effect. They found that the amplification factor can be maximized by tuning the gate temperature to its critical temperature. \cite{prod2016optimized}

\begin{figure}[b]
\includegraphics[width=0.5\textwidth]{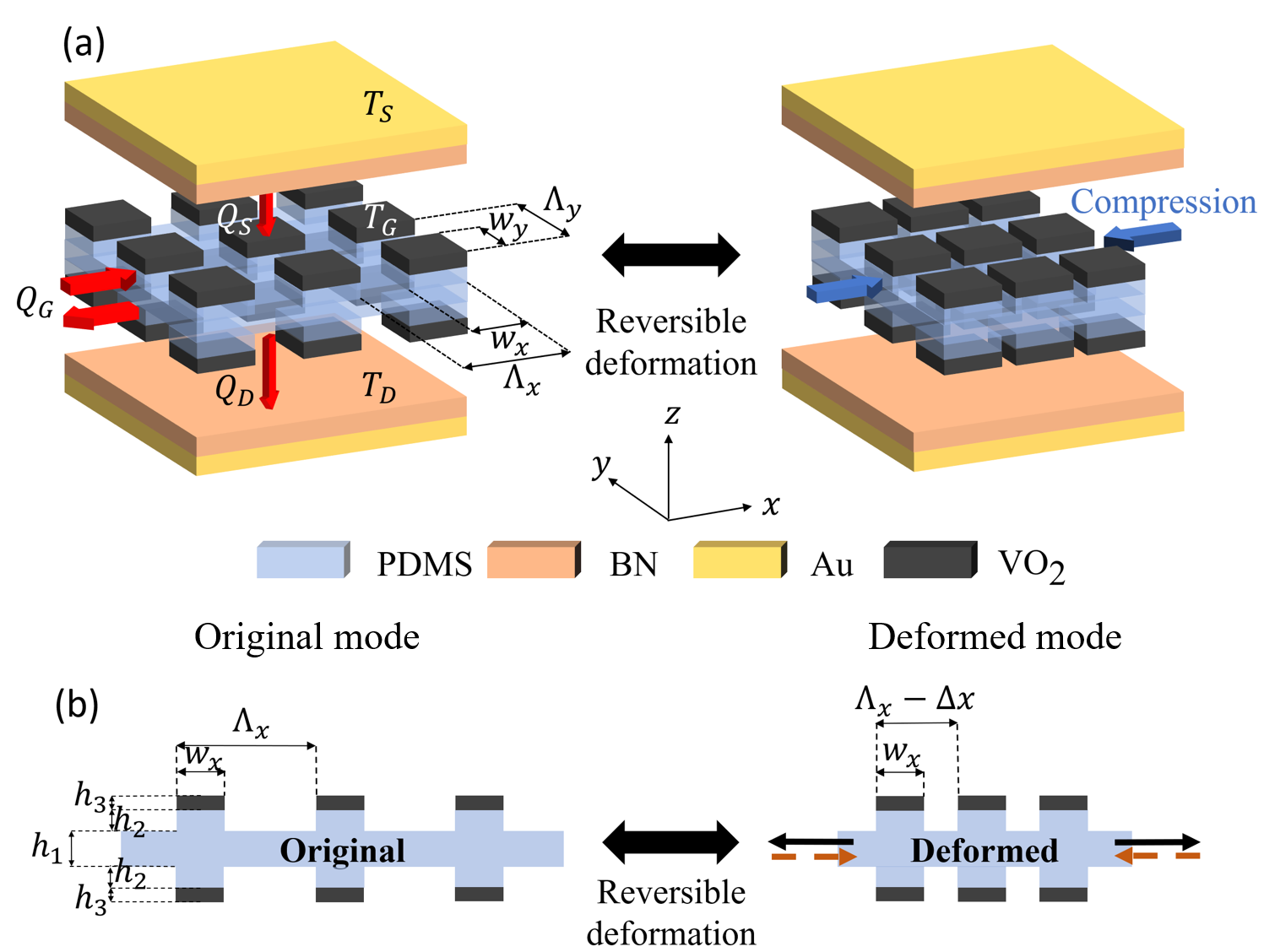}
\caption{\label{fig:fig1} (a) A reconfigurable non-contact thermal transistor based on near-field radiative heat transfer. From top to bottom, the three terminals are the source, the gate and the drain, analogous to the field-effect transistor. The structure of the gate terminal is varied when being subjected to external force (tension or compression). (b) The cross section of the gate terminal in $x-z$ plane. The thickness of the substrate $h_1$, the PDMS grating $h_2$, and the VO$_2$ grating $h_3$ are 120 nm, 100 nm, and 260 nm, respectively.}
\end{figure}

The phase-transition materials have been widely used in the thermal rectification. Among them, the most popular one is vanadium dioxide (VO$_2$) with the critical temperature of 341 K, which can be further tuned by the inclusion of tungsten. \cite{ito2014experimental} VO$_2$ is in anisotropic insulator state while below its critical temperature; otherwise, it is in metallic state. Therefore, the VO$_2$ is a potential candidate for radiative thermal rectification because of this temperature-dependent dielectric property. Ito \emph{et al.} experimentally investigated the far-field radiative thermal rectifier utilizing VO$_2$ and tuned its operating temperature by doping tungsten.\cite{ito2014experimental} Yang \emph{et al.} collected numerical results of thermal rectification based on VO$_2$ and discussed the influence of film thickness.\cite{yang2013radiation,yang2015vacuum} Joulain \emph{et al.} showed that phase change materials can perform modulation as well as amplification of radiative thermal flux. \cite{joulain2015modulation} The phase-transition materials can be included in the rectifier in several forms, such as bulk, thin film, grating, and nanoparticle. Ghanekar \emph{et al.} demonstrated the effects of VO$_2$ inclusions as the form of bulk, thin film and grating. \cite{ghanekar2016high} It indicates that the grating has way better performance in thermal rectification than other inclusion forms. In this work, VO$_2$ is included in the transistor in the form of a reconfigurable two-dimensional grating. In general, the two-dimensional grating will simplify to one-dimensional grating if either one of the filling ratios equals one. 

Dynamic tuning of the nanoscale radiative thermal transport is another focus addressed in this work. Gu \emph{et al.} proposed a new way to modulate the gate temperature by changing the position of the gate, which can be carried out with the piezoelectric motors experimentally. \cite{gu2015thermal} Besides, by changing the configuration of a thermal device while mechanical strain is applied, the thermal transport can also be modulated. Biehs \emph{et al.} presented a theoretical study showing that the twisting angle between two grating structures can modulate the net heat flux up to 90\% at room temperature.\cite{biehs2011modulation} Ghanekar \emph{et al.} proposed a near-field thermal modulator exhibiting sensitivity to mechanical strain. \cite{ghanekar2018strain} Liu \emph{et al.} demonstrated that a non-contact thermal modulator based on the mechanical rotation can achieve a modulation contrast greater than 5.\cite{liu2017pattern} Though few studies about thermal rectifiers driven by mechanical force have been conducted, reconfigurable meta-materials with manipulatable electromagnetic properties have drawn a lot of attention, and the related studies shed light on further research on nano-scale radiative thermal rectification. \cite{lee2012reversibly,pryce2010highly,zheludev2016reconfigurable} In this study, we use the reconfigurable periodic structures made of soft material polydimethylsiloxane (PDMS), which is easy to be deformed while subjected to external force, like tension or compression. This will result in distinct radiative heat fluxes due to different controllable filling ratios. The amplification factor of this thermal transistor can be dynamically tuned.

The proposed near-field thermal transistor with three parallel terminals is schematically plotted in Fig. \ref{fig:fig1}(a). From top to bottom, the three terminals are the source, the gate, and the drain. Both the source and drain terminals are composed of 1 $\mu$m BN layer on top of 1 $\mu$m gold layer. The gate terminal is composed of a two-dimensional grating with filling ratio $f_x=w_x/\Lambda_x$ and $f_y=w_y/\Lambda_y$ in $x$ and $y$ direction. $w_x$ and $w_y$ ($\Lambda_x$ and $\Lambda_y$) are the widths (periods) of the periodic structure in $x$ and $y$ direction. For the sake of clarity, the gate structure is displayed as the cross section in $x-z$ plane shown in Fig. \ref{fig:fig1}(b). The substrate is a PDMS film with thickness $h_1$, accompanied by two VO$_2$ rectangular gratings stacking on two PDMS gratings on both sides. As shown in Fig. \ref{fig:fig1}(b), the structure is symmetric in $x-z$ plane. The distance between the two adjacent terminals is kept at 100 nm, smaller than the thermal wavelength. In this design, the source temperature $T_S$ and the drain temperature $T_D$ are fixed at 371 K and 311 K, respectively. The temperature of the gate is controlled by applying a certain amount of heat $Q_G$ added to or removed from the gate so that the source heat flux $Q_S$ and the drain heat flux $Q_D$ can be tuned. The device thus behaves as a transistor: a small change $\Delta Q_G$ will lead to a big change $\Delta Q_S$ and $\Delta Q_D$. When compressing the grating, it is found that the amplification factor is dynamically tuned because of the change in the gate-terminal structure. In our work, the heights $h_1$ and $h_2$ are assumed to be constant during the deformation. Also, the mechanical deformation can be reversed if a tension is applied. More details will be discussed with the results.

The heat flux between the planar surfaces of the thermal transistor is obtained through the the dyadic Green's function formalism: \cite{narayanaswamy2014green}
\begin{equation}
Q=\int_{0}^{\infty} \frac{d \omega}{2 \pi}\left[\Theta\left(\omega, T_{1}\right)-\Theta\left(\omega, T_{2}\right)\right] \int_{0}^{\infty} \frac{k_{\parallel} d k_{\parallel}}{2 \pi} Z \left(\omega, k_{\parallel}\right)
\end{equation}
where $T_1$ and $T_2$ are the temperatures of the two surfaces, respectively. $\Theta(\omega, T)=(\hbar \omega / 2) \operatorname{coth}\left(\hbar \omega / 2 k_{B} T\right)$ is the energy of the harmonic oscillator. $\int_{0}^{\infty} \frac{k_{\parallel} d k_{\parallel}}{2 \pi} Z\left(\omega, k_{\parallel}\right)$ is known as the spectral transmissivity in radiative transfer between media 1 and 2 with gap $L$,
where $k_{\parallel}$ is the parallel component of wavevector and $Z\left(\omega, k_{\parallel}\right)$ is known as the energy transmission coefficient.

When the temperature of VO$_2$ is lower than its critical temperature (341 K), it is in the anisotropic insulator state. We can use the classical oscillator formula $\varepsilon(\omega)=\varepsilon_{\infty}+\sum_{i=1}^{N} \frac{S_{i} \omega_{i}^{2}}{\omega_{i}^{2}-j \gamma_{i} \omega-\omega^{2}}$ to determine  dielectric functions of the ordinary mode $\varepsilon_{O}$ and the extraordinary mode $\varepsilon_{E}$. Here, $\varepsilon_{\infty}$ is high-frequency constant, $\omega_{i}$ is phonon frequency, $\gamma_{i}$ is scattering rate, and $S_{i}$ is oscillator strength.
As VO$_2$ is in the completely isotropic metallic state, the Drude model is used to calculate the dielectric function $\varepsilon(\omega)=\frac{-\omega_{p}^{2} \varepsilon_{\infty}}{\omega^{2}-j \omega \Gamma}$. Here, $\omega_{p}$ is plasma frequency and $\Gamma$ is collision frequency.
 The experimental values for calculation are given in the work of Barker \emph{et al.}.\cite{barker1966infrared} 
 The dielectric functions of gold, PDMS and BN can be found in these work. \cite{johnson1972optical,querry1987optical,palik1998handbook}

For asymmetric two-dimensional grating consisting of two materials with dielectric function $\epsilon_A$ and $\epsilon_B$, the filling ratios of the inclusion material B in the host material A are $f_x$ and $f_y$ (the filling ratios can have different values in $x$ and $y$ direction). The effective refractive index of medium is derived by including another filling ratio in the original calculation methods for symmetric two-dimensional grating structure: \cite{chen2013polarity,brauer1994design}
\begin{equation}
n_{2-D}=\left[\bar{n}+\hat{n}_x+\hat{n}_y+\check{n}_{x}+\check{n}_{y}\right] / 5
\end{equation}

Each term can be obatined as follows: $\bar{n}=\left(1-f_xf_y\right) n_{A}+f_xf_y n_{B}$, $\hat{\varepsilon}_{x}=(1-f_y) \varepsilon_{A}+f_y \varepsilon_{x,\perp}$, $1 / \check{\varepsilon}_{x}=(1-f_x) / \varepsilon_{A}+f_x / \varepsilon_{y,\|}$ , 
where $\varepsilon_{\|}$ and $\varepsilon_{\perp}$ are obtained by $\varepsilon_{(x,y),\|}=(1-f_{(x,y)}) \varepsilon_{A}+f_{(x,y)} \varepsilon_{B}$ and $1 / \varepsilon_{(x,y),\perp}=(1-f_{(x,y)}) / \varepsilon_{A}+f_{(x,y)} / \varepsilon_{B}$. $\hat{\varepsilon}_{y}$ and $\check{\varepsilon}_{y}$ are obtained by interchanging the $x$ and $y$ subscripts.

\begin{figure}[b]
\includegraphics[width=0.45\textwidth]{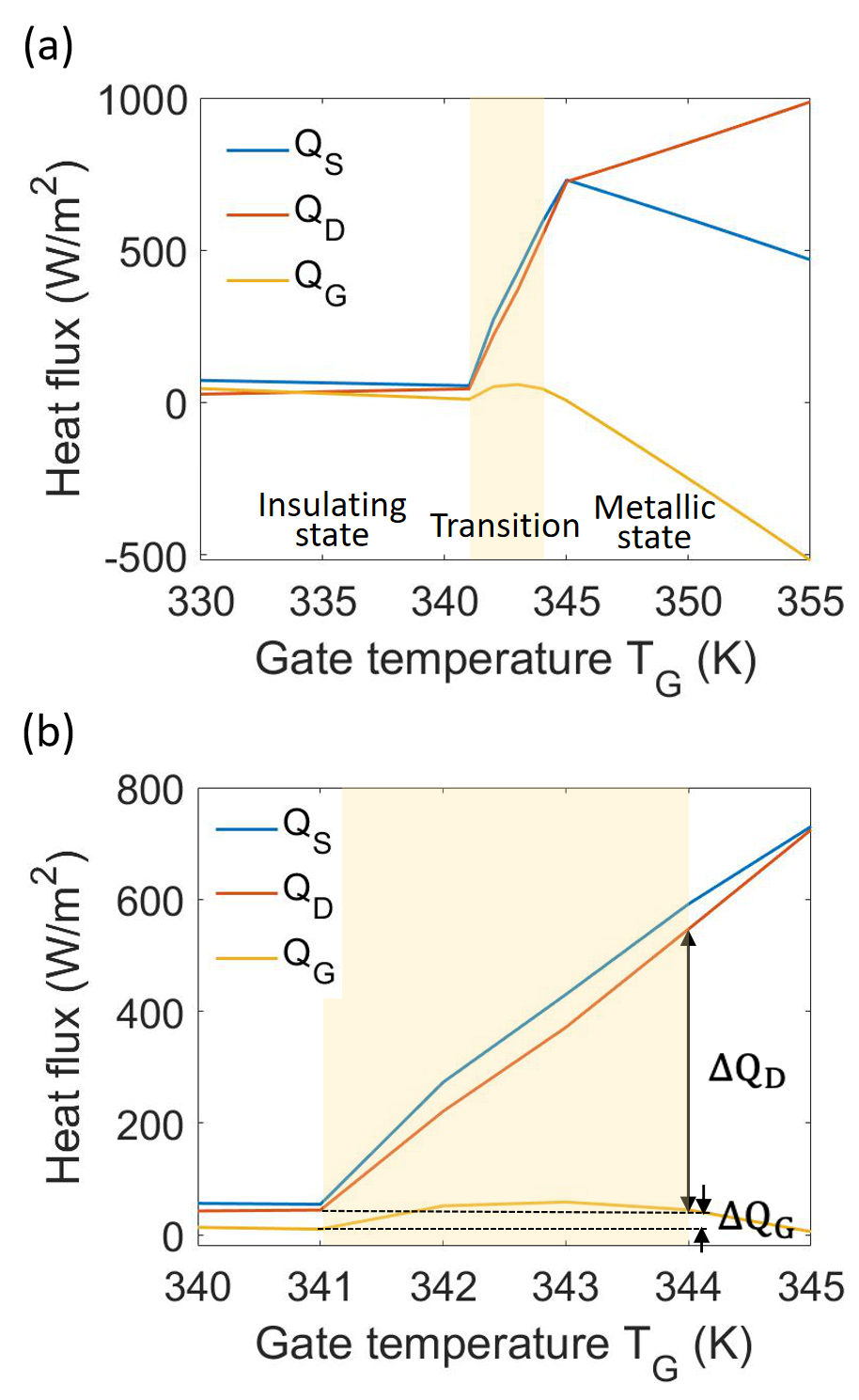}
\caption{\label{fig:fig2} (a) The variation of heat fluxes $Q_S$, $Q_D$ and $Q_G$ depending on the gate temperature $T_G$. $T_G$ varies from the 330 K to 355 K. The region from 341 K to 344 K is the phase-transition region, which is highlighted by yellow. The filling ratios $f_x$ and $f_y$ are 0.5 and 0.6, respectively. (b) Zoom in on Fig. \ref{fig:fig2}(a) to the temperature region from 340 K to 345 K.}
\end{figure}

Although the critical temperature of VO$_2$ is 341 K, the phase change does not happen instantaneously once the temperature reaches 341 K. \cite{yang2015vacuum} Here, it is reasonably assumed the phase-transition region is from 341 K to 344 K, where the dielectric function of VO$_2$ is modeled using  the effective medium theory. As the temperature goes up, metallic puddles are formed with increasing volume fraction to 1 at 344 K. \cite{qazilbash2009infrared} Bruggeman approximation can also be applied. \cite{ghanekar2018near} One of the most important features of the thermal transistor is its capacity to amplify the heat flux received by the drain through a small change in $Q_G$. In Fig. \ref{fig:fig2}, the filling ratios $f_x$ and $f_y$ are 0.5 and 0.6. When VO$_2$ is in completely insulating state ($T_G<341$ K) or metallic state ($T_G>345$ K), all the three heat fluxes change with the gate temperature in a linear trend, $\alpha\approx0.5$, and no amplification is observed. This is because, in these regions, the dielectric properties of VO$_2$ are independent of temperature, and, as $T_G$ being varied , the slopes of $Q_S - T_G$ and $Q_D - T_G$ are opposite to each other, \emph{i.e.}, $\partial Q_S/\partial T_G=-\partial Q_D/\partial T_G$. \cite{ben2014near} The transistor-like behavior of this device in the transition region (yellow section) is clearly shown in Fig. \ref{fig:fig2}(b). $Q_S$ and $Q_D$ rise steeply with a small variation in $Q_G$. In terms of the original definition of the amplification factor, $\alpha$ will go up to unreasonable high values if $Q_G$ is around its maximum. Therefore, $\alpha$ is alternatively defined as $|\Delta Q_D/\Delta Q_G|$. $\Delta Q$ is calculated using the difference between the heat fluxes at 341 K and 344 K, which are the boundaries of the phase transition region. The difference between the heat fluxes across the entire phase change is a good approximation of the differential in the original definition of $\alpha$.

The equilibrium temperature of the gate $T_e$ can be derived from Stefan-Boltzmann's law and the principle of energy conservation, $T_{e}=\sqrt[4]{\left(T_{S}^{4}+T_{D}^{4}\right) / 2}$. It is the temperature when $Q_G$ equals to zero. Compared with the previous work, \cite{ghanekar2018near} as $T_S$ and $T_D$ being chosen as 371 K and 321 K, the equilibrium temperature $T_e$ is 348.7 K, which is larger than 344 K and is not in the phase-transition region (341 K $-$ 344 K). The device reaches the equilibrium when VO$_2$ is in completely metallic state. When $T_S$ and $T_D$ are 361 K and 311 K, respectively, the equilibrium temperature $T_e$ is 338.8 K, which is smaller than 341 K, and VO$_2$ is in insulating state. In our case, $T_S$, $T_D$, $T_e$ are 371 K, 311 K, and 344.9 K, respectively. $T_e$ is closer to the right boundary of the phase-transition region, and consequently, the change of $Q_G$ becomes much smaller. In other words, with appropriate source and drain temperatures, a high-performance transistor with a large amplification factor is obtained.

\begin{figure}
\includegraphics[width=0.45\textwidth]{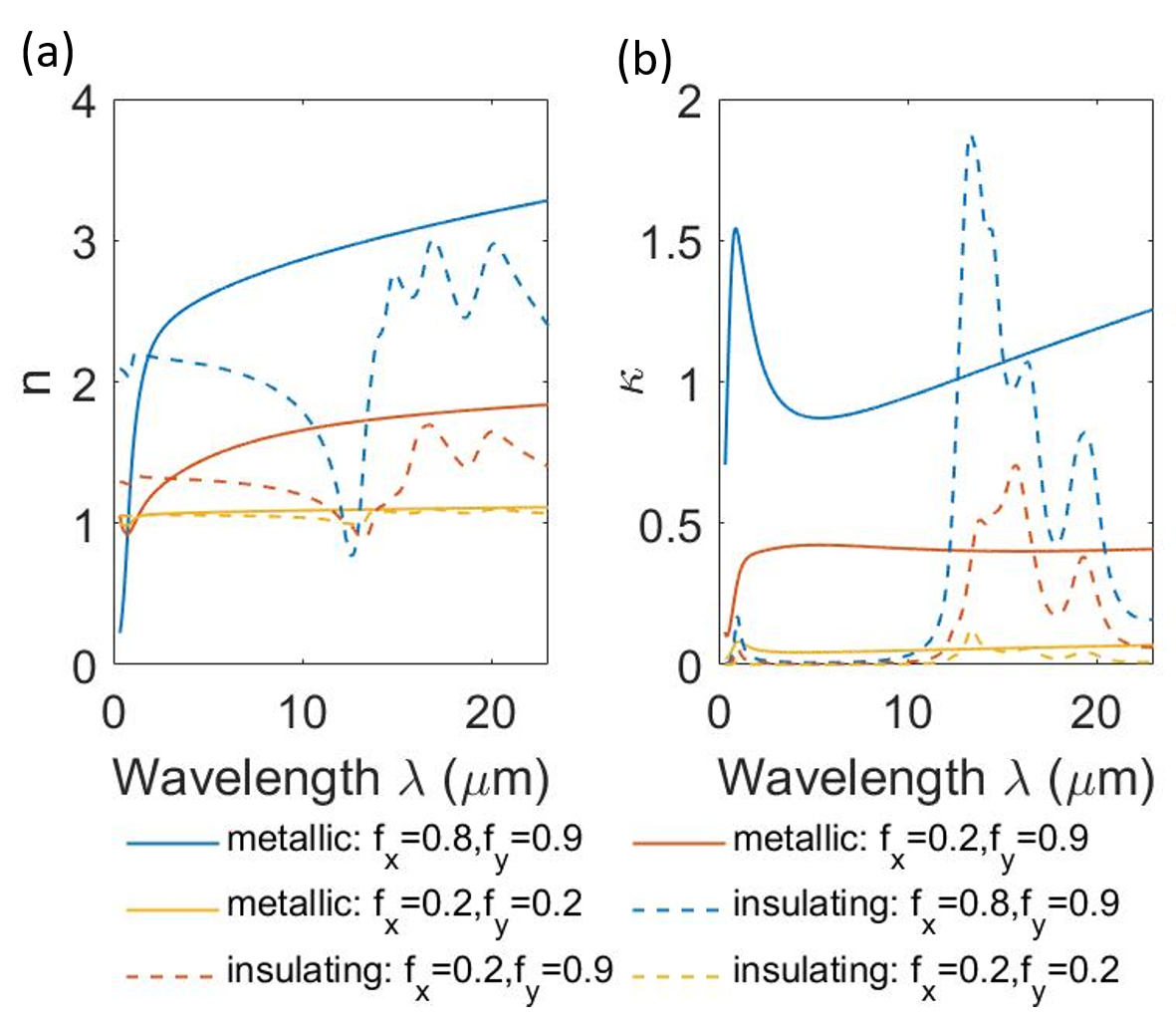}
\caption{\label{fig:fig3} (a) Refractive index $n$ and (b) extinction coefficient $\kappa$ of VO$_2$ grating for metallic and insulating states in three different filling ratio patterns: (1) $f_x=0.8, f_y=0.9$, (2) $f_x=0.2, f_y=0.9$, and (3) $f_x=0.2, f_y=0.2$.}
\end{figure}

The refractive index $n$ and extinction coefficient $\kappa$ are the real and imaginary parts of the square root of dielectric function, $\sqrt{\epsilon}=n+i\kappa$. As shown in Fig. \ref{fig:fig3}, $n$ and $\kappa$ of the VO$_2$ grating are plotted at both metallic and insulating states with three different filling ratio patterns. When the filling ratios of VO$_2$ decrease, both $n$ and $\kappa$ will be smaller, showing a smaller influence from VO$_2$. The insulating VO$_2$ supports surface phonon polaritons (SPhPs) beyond the wavelengths of 10 $\mu$m, while the metallic VO$_2$ supports surface plasmon polaritons (SPPs). 
SPPs/SPhPs occur at the interfaces across a nanoscale gap, and they exist between planer surfaces, curvature structures, or the ones with surface patterns such as periodic grating structures. \cite{van2011phonon,shen2009surface,zheng2017excitation,zayats2005nano} They exhibit a large local-field enhancement to radiative heat flux near the interfaces. In this work, since we adopt the reconfigurable grating induced by the soft material PDMS, the deformation of the grating can be precisely controlled by an external force, resulting in the change in the modes of SPhPs/SPPs and the coupling effects of the surface waves along the interfaces, particularly in the phase-transition temperature range. Simultaneously, surface waves are constrained between the boundaries of two adjacent pillars along the periodic interface, and the modes of SPhPs/SPPs become variable depending on the mechanical deformation.

Under mechanical tension and compression, the surface phonon polaritons along the interface between the insulating VO$_2$ periodic structure and the vacuum gap are affected dramatically, which play a significant role in modulating the near-field radiative thermal transport. It yields a dynamic control of amplification in a non-contact thermal transistor.

\begin{figure}
\includegraphics[width=0.45\textwidth]{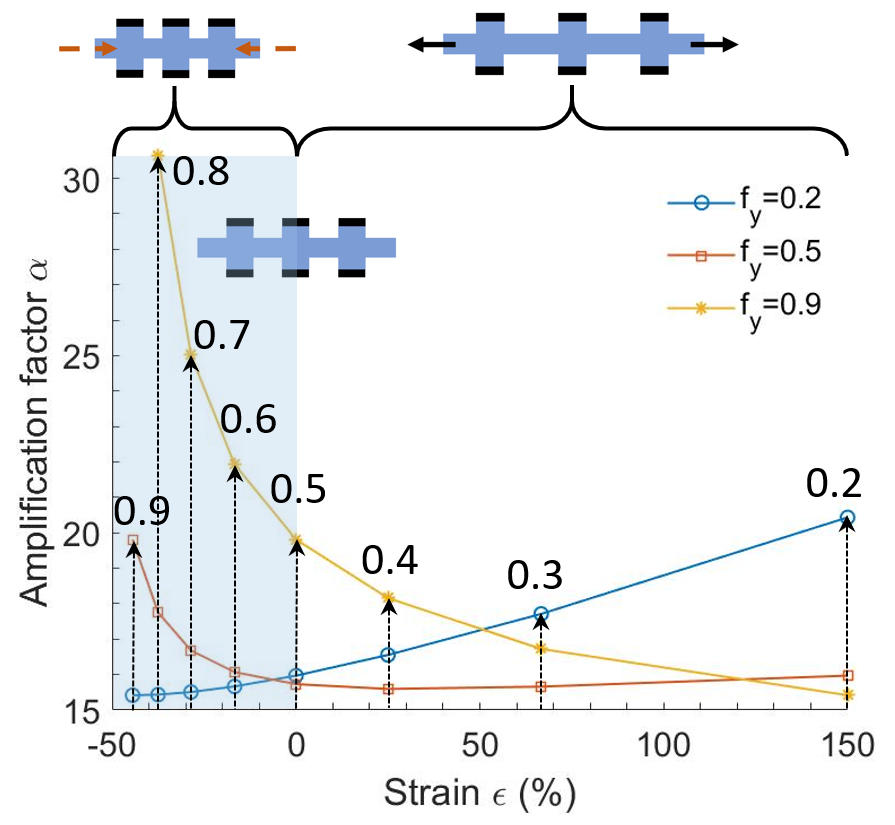}
\caption{\label{fig:fig4} Dynamic tuning of amplification factor $\alpha$ due to the compression and tension on the gate terminal. The zero strain point is set as $f_x=0.5$. The strain is positive with $f_x$ smaller than 0.5 when tensile force is applied, while it is negative with $f_x$ larger than 0.5 when compressive force is applied (blue background). $f_y$ is fixed and $f_x$ varies from 0.2 to 0.9.}
\end{figure}

The large amplification factor $\alpha$ is definitely an expected goal for a transistor, while a modulable amplification factor is another feature of the proposed device. Figure \ref{fig:fig4} displays how $\alpha$ changes when filling ratio in $x$ direction, $f_x$, varies from 0.2 to 0.9, and meanwhile filling ratio in $y$ direction, $f_y$, is fixed at three distinct values, 0.2, 0.5 and 0.9. The strain is defined as the change in the period divided by the original grating period before any compression or tension ($\Delta x/\Lambda_x$). Here, the zero strain point is set as $f_x=0.5$. When the gate terminal is subjected to a tensile force, $f_x$ will decrease and the strain is positive. On the other hand, as the structure is under compression, $f_x$ will go up and the strain is negative. The blue curve ($f_y=0.2$) exhibits a slightly increasing trend for $\alpha$ as the filling ratio decreases in the $x$ direction. However, for the orange ($f_y=0.5$) and yellow curves ($f_y=0.9$), no such trend appears at small $f_x$. For these two curves, the increase of $\alpha$ occurs at larger $f_x$ and the curve for $f_y=0.9$ shows a steeper rising trend. Hence, the tension will only contribute to modulating the amplification factor when $f_y$ is very small, and so will the compression only when $f_y$ is relatively large. It is also worth noticing that the status of $f_x$ and $f_y$ are the same. Here, we are discussing the case of varying $f_x$ while keeping $f_y$ constant. However, both filling ratios $f_x$ and $f_y$ play an important role in the modulable amplification factor. So here we define a new parameter, filling area, to evaluate the performance of a two-dimensional grating based non-contract thermal transistor. The filling area is defined as $f_a=f_x
\times f_y$. Only when $f_a\leq 0.09$ or $f_a\geq 0.35$, the modulation of the amplification factor $\alpha$ is observable.

Analogous to the electric transistor, which is already a mature device in microelectronics, the thermal transistor is a promising device that can act as a switch, a modulator, or an amplifier. A dynamic-tuning non-contact thermal transistor with grating nanostructure is studied in this work. It is composed of the source, the gate, and the drain terminals. The gate terminal is made of two VO$_2$ rectangular
gratings stacking on two PDMS gratings on both sides of a PDMS substrate. The source and the drain thermal flow can be drastically enhanced due to the small variation of the gate thermal flow. This transistor-like amplification function occurs only in the phase-transition region of VO$_2$, which is assumed to be between 341 K and 345 K in this design, while no amplification is observed beyong this temperature region. Besides, by applying a compressive or tensile force on the stretchable gate terminal, the amplification factor can be dynamically tuned. The separations between the source and the gate, the gate and the drain are fixed at 100 nm. The grating height also remains unchanged. As we keep changing the filling ratios in both $x$ and $y$ directions, the amplification factor varies from 15.4 to 30.6. A further study can be carried out in the future, such as changing the fixed temperatures for the source and the drain terminals. The two separation distances can be different to form an asymmetric structure. Alternative phase-transition and host materials and other nanostructures of the gate terminal would also yield different outcomes. This work sheds light on the high-performance thermal transistors and motivates potential applications in dynamic tuning of nanoscale thermal transport due to mechanical deformations.

~\\
\indent This project was supported by the National Science Foundation through Grant No. 1941743.

~\\
\indent The data that support the findings of this study are available from the corresponding author
upon reasonable request.
\nocite{*}
\bibliography{aipsamp}

\end{document}